# Speedup of Micromagnetic Simulations with C++ AMP On Graphics Processing Units

Ru Zhu[1]

Graceland University, Lamoni, Iowa 50140 USA

**Abstract**

A finite-difference Micromagnetic solver is presented utilizing the C++ Accelerated Massive Parallelism (C++ AMP). The high speed performance of a single Graphics Processing Unit (GPU) is demonstrated compared to a typical CPU-based solver. The speed-up of GPU to CPU is shown to be greater than 100 for problems with larger sizes. This solver is based on C++ AMP and can run on GPUs from various hardware vendors, such as NVIDIA, AMD and Intel, regardless of whether it is dedicated or integrated graphics processor.

**1. Introduction**

Micromagnetic simulations are important tools to study magnetic dynamics and design new magnetic devices. Micromagnetic solvers running on Central Processing Unit (CPU) such as OOMMF [1] and magpar [2] have been widely used in magnetism research. Micromagnetic simulations of complex magnetic structures require fine geometrical discretization, and are time consuming.

There has been growing research work on applying general purpose Graphics Processing Units (GPU) in the fields of Micromagnetics, such as MuMax, FastMag and GPMagnet [3] – [7]. Due to the high computing power of GPU units, these works have achieved considerable speed-ups as compared to previous CPU based implementations. On the other hand, general purpose GPU units are cheap, most costing less than $1000. Therefore complex micromagnetic simulations can be done at much lower cost.

However, these implementations are exclusively based on NVIDA's parallel computing platform Compute Unified Device Architecture (CUDA) and their applications are limited to NVIDIA GPUs. In 2012, Microsoft released its own parallel programming library named C++ AMP which is an open specification and is hardware platform independent [8]. Software based on C++ AMP can run on virtually all latest GPUs, including those from NVIDIA, AMD and Intel. The purpose of this work then is to implement a cross-platform micromagnetic solver for solving the Landau-Lifshitz-Gilbert (LLG) equation.

Section 2 discusses the formulation of the magnetic dynamics and LLG equation, especially the formulas used to calculate the demagnetization field. Section 3 describes the software implementation of the micromagnetic solver. Section 4 presents the performance of this micromagnetic solver at various problem sizes, and compares it with a popular CPU-based micromagnetic solver.

---

[1] Email address: zhu@graceland.edu.

## 2. Principle

Consider a magnetization vector $\vec{M} = (M_x, M_y, M_z)$ in a computational cell that belongs to the computing region. Its saturation magnetization $M_s = \sqrt{M_x^2 + M_y^2 + M_z^2}$. The magnetic energy density related to this vector can be written down as

$$\varepsilon = A[(\nabla \frac{M_x}{M_s})^2 + (\nabla \frac{M_x}{M_s})^2 + (\nabla \frac{M_x}{M_s})^2] + K_u \frac{(M_y^2 + M_z^2)}{M_s^2} \\ - \frac{1}{2}(\mu_0 \vec{H}_{demag} \vec{M}) - (\mu_0 \vec{H}_{extern} \vec{M}) \tag{1}$$

The right hand side of (1) consists of the exchange, anisotropy, demagnetization and Zeeman energy densities, where $A$ is the material exchange constant, $K_u$ is the uniaxial anisotropy constant, $\mu_0$ is the vacuum permeability, $H_{demag}$ is the demagnetization field and $H_{extern}$ is the external field. The anisotropy energy is assumed to be uniaxial with an easy axis on the $x$ direction.

The change of magnetization vector is caused by the effective magnetic field $H_{eff}$ derived from the magnetic energy density:

$$\vec{H}_{eff} = -\frac{\delta \varepsilon}{\delta \vec{M}} = \vec{H}_{exch} + \vec{H}_{anis} + \vec{H}_{demag} + \vec{H}_{extern} \tag{2}$$

where $\frac{\delta \varepsilon}{\delta \vec{M}}$ gives the functional derivative of $\varepsilon$ with respect to $\vec{M}$. In (2) the exchange field and anisotropy field are represented by $\vec{H}_{exch}$ and $\vec{H}_{anis}$ respectively.

According to (1) and (2),

$$H_{exch,x} = \frac{2A}{M_s^2} \nabla^2 M_x. \tag{3}$$

To derive the exchange field $H_{exch}$ we need to discretize the computing region properly and consider the magnetizations of neighboring computational cells. The entire computing region is divided into $n_x \times n_y \times n_z$ cells, each cell with an equal volume of $\delta x \delta y \delta z$. The cells are labeled with indices

$0 \leq i \leq n_x - 1,$
$0 \leq j \leq n_y - 1,$
$0 \leq k \leq n_z - 1.$

Note that *i*, *j* and *k* are zero-indexed to follow the convention of C++ programming language.

According to (3), the Cartesian components of the effective field can be expressed as

$$H_{exch,x} = \frac{2A}{M_s^2} \{ \frac{M_x(i+1,j,k) - 2M_x(i,j,k) + M_x(i-1,j,k)}{\delta x^2}$$
$$+ \frac{M_x(i,j+1,k) - 2M_x(i,j,k) + M_x(i,j-1,k)}{\delta y^2} \quad (4)$$
$$+ \frac{M_x(i,j,k+1) - 2M_x(i,j,k) + M_x(i,j,k-1)}{\delta z^2} \}.$$

Other components of $\vec{H}_{exch}$ can be obtained by replacing *x* with *y* or *z* in (4).

According to (1) and (2),

$$H_{anis,x} = -\frac{2K_u}{\mu_0 M_s^2} M_x. \quad (5)$$

The LLG equation in the low damping limit is [9]

$$\frac{d\vec{M}}{dt} = -\frac{\gamma}{1+\alpha^2}(\vec{M} \times \mu_0 H_{eff}) - \frac{\alpha\gamma}{(1+\alpha^2)M_s}[\vec{M} \times (\vec{M} \times \mu_0 \vec{H}_{eff})] \quad (6)$$

where *α* is the damping constant, and *γ* is the gyromagnetic ratio.

To speed up the micromagnetic simulation, it is necessary to decrease per-step simulation time, most of which is consumed by the calculation of the demagnetization field. The brute force calculation of demagnetization field is known to be proportional to the square of the number *N* of the computational cells [10]. However, this calculation can be accelerated by taking advantage of the discrete convolution theorem and the fast Fourier transform (FFT) [11].

Consider a rectangular magnetic sample with its edges parallel to Cartesian coordinate axes *x*, *y* and *z*. For simplicity, we will start from one-dimensional case. Divide the sample into $n_x$ cells on the *x* direction, and label each cell with an index *i*, $i = 0, \ldots, n_x - 1$. Then the demagnetization field applied on cell *i* is

$$H_{demag}(i) = \sum_{l=0}^{n_x - 1} M(l) K(l-i) \quad (7)$$

Where *K(l-i)* is the demagnetization tensor giving the contribution of source cell *M(l)* to observation cell *i*. Since the tensor *K* is solely determined by the difference of *l* and *i*, and it is obvious that $H_{demag}$ is a convolution of *K* (demagnetization tensor) and *M* (magnetization).

To avoid the effect of cyclic convolution, we have to do zero-padding to original data so that fast Fourier transform can be performed. It is also necessary to achieve correct output in real space after performing the inverse FFT as described in step (5) below. Here is how the zero-padding algorithm can be implemented [11]:

(1) Set $M(i)$, $i = n_x, \ldots, 2n_x - 1$ to zero while keeping the original M(i) $i = 0, \ldots, n_x - 1$ intact.

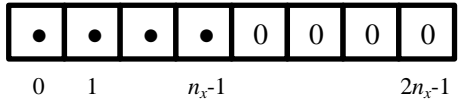

(2) Set $K(2n_x - 1) = 0$ while keeping the original $K(i)$, $i = -(n_x - 1), \ldots, 0, \ldots, n_x - 1$ intact.

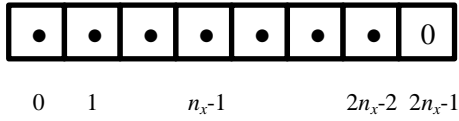

(3) Carry out Discrete Fourier Transform (DFT) to the padded $M$ and $K$ data with FFT algorithm:

$$\tilde{M} = FFT(M)$$
$$\tilde{K} = FFT(K) \tag{8}$$

(4) According to the DFT theorem, the DFT of right hand side of (7) is member-wise dot product of DFTs of $M$ and $K$:

$$\tilde{H}_{demag}(i) = \tilde{M}(i) \cdot \tilde{K}(i) \tag{9}$$

(5) carry out inverse DFT on $\tilde{H}_{demag}$ to get $H_{demag}$:

$$H_{demag} = FFT^{-1}(\tilde{H}_{demag}). \tag{10}$$

Keep in mind that due to zero-padding $\tilde{H}_{demag}$ has redundant data and is twice as large as $H_{demag}$, and as a result we need to discard the latter half of the data from $FFT^{-1}(\tilde{H}_{demag})$.

In three-dimensional space a rectangular magnetic sample can be divided into $n_x \times n_y \times n_z$ cells. After zero padding the input data size increases to $2n_x \times 2n_y \times 2n_z$, as demonstrated by Fig. 1.

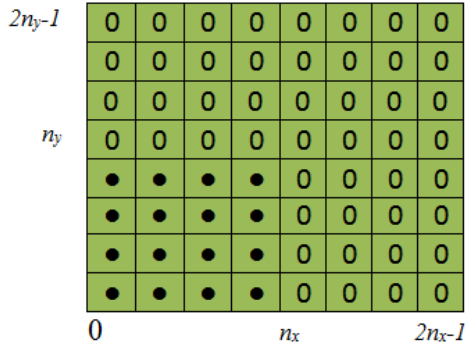

Fig. 1 A cross-sectional view of a rectangular magnetic sample after zero-padding.

In the case of a finite three-dimensional computing problem, the demagnetization field can be calculated as

$$H_{demag}(i,j,k) = \sum_{l=0}^{n_x-1}\sum_{m=0}^{n_y-1}\sum_{n=0}^{n_z-1} M(l,m,n)K(l-i,m-j,n-k) \tag{11}$$

or

$$\begin{aligned}H_{demag,x}(i,j,k) = \sum_{l=0}^{n_x-1}\sum_{m=0}^{n_y-1}\sum_{n=0}^{n_z-1} &\{M_x(l,m,n)K_{xx}(l-i,m-j,n-k)\\ &+ M_y(l,m,n)K_{xy}(l-i,m-j,n-k)\\ &+ M_z(l,m,n)K_{xz}(l-i,m-j,n-k)\}\end{aligned} \tag{12}$$

Other components of $H_{demag}$ can be obtained by replacing x with y or z in (12).

By applying DFT theorem to both sides of the equation, we can get

$$\begin{aligned}\tilde{H}_{demag,x}(i,j,k) &= \tilde{M}_x(i,j,k)\cdot \tilde{K}_{xx}(i,j,k)\\ &+ \tilde{M}_y(i,j,k)\cdot \tilde{K}_{xy}(i,j,k)\\ &+ \tilde{M}_z(i,j,k)\cdot \tilde{K}_{xz}(i,j,k)\end{aligned} \tag{13}$$

Finally, the demagnetization field $H_{demag}$ can be obtained by taking the inverse FFT of $\tilde{H}_{demag}$, as described by (10).

## 3. Implementation

GPUs have intrinsically different hardware architecture from CPUs, notably for its large number of Arithmetic Logic Units (ALU) that was initially designed for graphics rendering but now also used for general purpose computing. Since GPU is specialized for computing-intensive, highly parallel computation, it is ideally suitable for micromagnetic simulations in which large number of computational cells can be processed in parallel. This idea is schematically illustrated by Fig. 2.

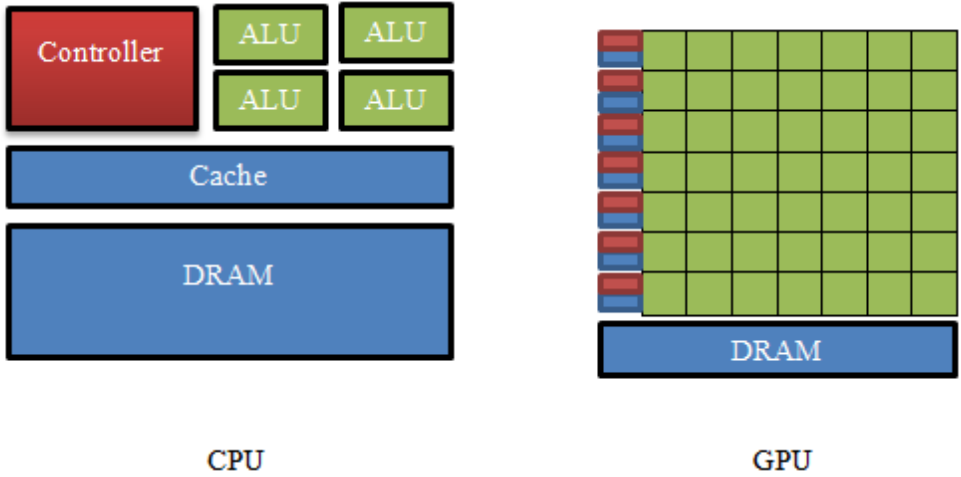

Fig. 2 A comparison between Hardware architectures of CPU and GPU. The GPU has more ALUs dedicated to data processing.

C++ AMP was implemented with High Level Shading Language (HLSL), which was initially designed for Microsoft's graphics library DirectX [8]. Compared with popular GPU programming languages such as CUDA, it is fully cross-platform, which means the programs written in C++ AMP can be migrated to another hardware vendor without any modification. Compared with other cross-platform GPU programming language such as Open Computing Language (OpenCL), it has much simplified Application Programming Interface (API), thus reducing the programming effort of programmers. Readers can refer to Append. A for a comparison between APIs of OpenCL and C++ AMP.

GPUs usually have their own memory, also known as graphic memory. The data I/O is very fast between its ALUs and its own memory (> 100 GB/s), compared to the I/O between GPU and CPU (about 10 GB/s). Therefore the bottleneck to boost GPU computing performance is the data transfer from CPU to GPU or vice versa. In the micromagnetic solver presented, the only data transfer between CPU and GPU takes place when the initial conditions of the computing region are specified and when the final data is calculated by GPU and transferred back to CPU for display. In this way the simulation speed can be maximized.

As mentioned before, the most time-consuming part of micromagnetic simulation is the calculation of demagnetization field. In each time step, the calculation requires three different phases:

a. Perform FFTs to magnetization components:

$$\tilde{M}_x = FFT(M_x),$$
$$\tilde{M}_y = FFT(M_y), \quad (14)$$
$$\tilde{M}_z = FFT(M_z).$$

b. Do member-wise product of result of (14) and FFT of demagnetization tensors:

$$\tilde{H}_{demag,x} = \tilde{M}_x \cdot \tilde{K}_{xx} + \tilde{M}_y \cdot \tilde{K}_{xy} + \tilde{M}_z \cdot \tilde{K}_{xz},$$
$$\tilde{H}_{demag,y} = \tilde{M}_x \cdot \tilde{K}_{xy} + \tilde{M}_y \cdot \tilde{K}_{yy} + \tilde{M}_z \cdot \tilde{K}_{yz}, \qquad (15)$$
$$\tilde{H}_{demag,z} = \tilde{M}_x \cdot \tilde{K}_{xz} + \tilde{M}_y \cdot \tilde{K}_{yz} + \tilde{M}_z \cdot \tilde{K}_{zz}.$$

c. Carry out inverse FFT of result of (15):

$$H_{demag,x} = FFT^{-1}(\tilde{H}_{demag,x}),$$
$$H_{demag,y} = FFT^{-1}(\tilde{H}_{demag,y}), \qquad (16)$$
$$H_{demag,z} = FFT^{-1}(\tilde{H}_{demag,z}).$$

In three-dimensional space, there are six FFTs to perform for each time step. The FFTs of demagnetization tensor $K$ have been carried out at the beginning of simulation and will not be taken later, since $K$ is constant.

A FFT library based on C++ AMP has been implemented before [12]. It is adapted to the calculation of demagnetization field in the micromagnetic solver. At the point of publication the FFT library can only handle single-precision floats so this solver is currently limited to single-precision computing.

## 4. Results

The micromagnetic standard problem 3 [13] was used to test the performance of this solver. A cubic magnetic particle is divided in to grids of $N \times N \times N$ and the minimum energy state is reach by applying the LLG equation to each computational cell. The relaxation process involves the magnetization dynamics under the influence of demagnetization field, exchange field and uniaxial anisotropy field. To benchmark the solver presented, a hardware system with Intel Xeon E5410 CPU and an AMD Radeon HD 7970 GHz Edition GPU was used. The GPU chipset was among the fastest on the consumer market but still cost less than $500. For comparison, the benchmark of CPU micromagnetic solver OOMMF is also presented, with data from the report of another research group who used an Intel i7-930 CPU [4]. Dimensions with powers of two are benchmarked to demonstrate the performance of solvers varying with problem size, as shown in table 1. However this magnetic solver can solve problems of any size limited by the graphic memory allocable by the GPU.

It is noticeable that at smaller problem sizes (N < 20) GPU solver is not significantly faster or even slower than CPU solver. This is caused by two factors. The first factor is that the data I/O overhead. The data transfer between GPU and CPU's main memory takes time. For a smaller problem size the calculation on GPU can be completed very soon, so in this case the computing power of GPU will not be fully utilized. For larger problems the data I/O time can be negligible when compared to the computing time. The second factor is the kernel launching overhead of GPU. This overhead is a constant regardless of the problem size, thus it is significant when the problem size is small.

**Table 1**. Per-step simulation time needed by CPU and GPU solvers for different 3D problem sizes ($N \times N \times N$) with the Euler algorithm. Numbers are in milliseconds.

| Size | CPU (ms) | GPU (ms) | speedup |
|---|---|---|---|
| $8^3$ | 0.8492 | 1.95 | ×0.43 |
| $16^3$ | 4.066 | 2.723 | ×1.5 |
| $32^3$ | 36.14 | 3.151 | ×11 |
| $64^3$ | 489.6 | 6.558 | ×74 |
| $128^3$ | 4487 | 26.34 | ×170 |

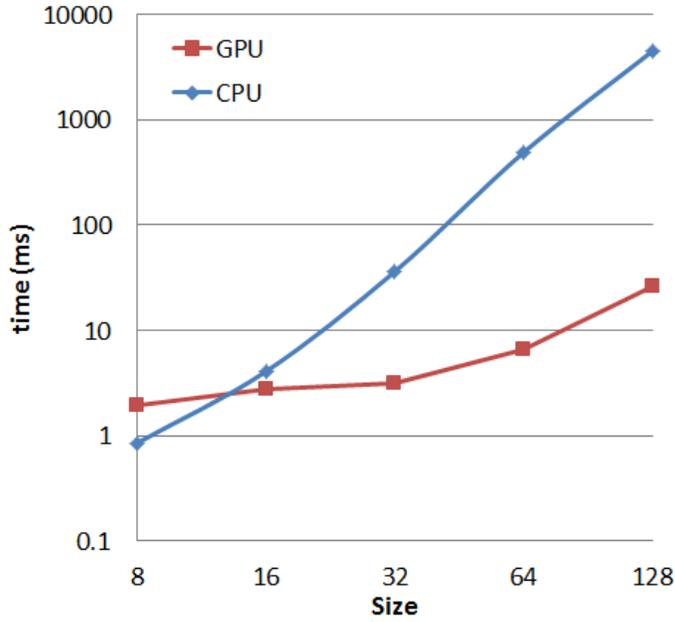

**Fig. 3**. Time need to carry out one time step at different 3D problem sizes $N \times N \times N$. The CPU time data is taken from report by [4].

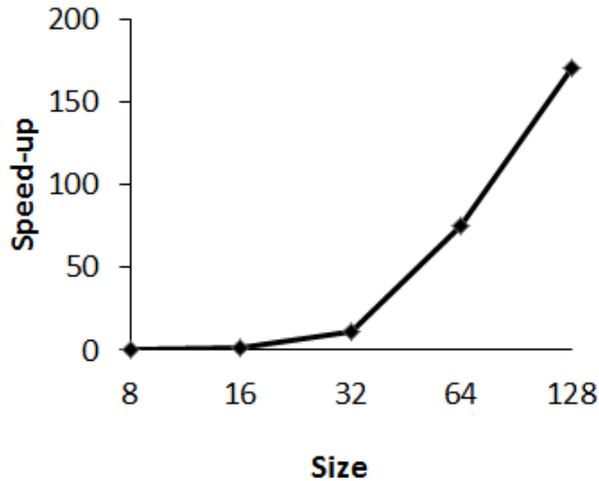

**Fig. 4**. Speed-up of GPU solver on an AMD Radeon HD 7970 GHz Edition compared to CPU solver OOMMF. The speed-up increases with problem size.

## 5. Summary

A GPU-based micromagnetic solver is presented to address the slow speed problem of large simulation problems. The speed boost relative to CPU simulations is significant at problem with large input sizes. This solver can not only run on expensive professional workstations but also economy personal laptops and both achieve considerable speed-ups.

## Acknowledgements

This work is supported by Graceland University professional development program. The author wishes to acknowledge Dr. Pieter B. Visscher of the University of Alabama, for helpful discussions of demagnetization field calculation.

## References

[1] Donahue, Michael Joseph, and Donald Gene Porter. OOMMF User's guide. US Department of Commerce, Technology Administration, National Institute of Standards and Technology, 1999.

[2] Scholz, Werner, et al. "Scalable parallel micromagnetic solvers for magnetic nanostructures." Computational Materials Science 28.2 (2003): 366-383.

[3] Kakay, Attila, Elmar Westphal, and Riccardo Hertel. "Speedup of FEM micromagnetic simulations with Graphics Processing Units." Magnetics, IEEE Transactions on 46.6 (2010): 2303-2306.

[4] Vansteenkiste, Arne, and Ben Van de Wiele. "MuMax: a new high-performance micromagnetic simulation tool." Journal of Magnetism and Magnetic Materials 323.21 (2011): 2585-2591.

[5] Chang, R., et al. "FastMag: Fast micromagnetic solver for complex magnetic structures." Journal of Applied Physics 109.7 (2011): 07D358.

[6] Li, Shaojing, Boris Livshitz, and Vitaliy Lomakin. "Graphics processing unit accelerated micromagnetic solver." Magnetics, IEEE Transactions on 46.6 (2010): 2373-2375.

[7] Lopez-Diaz, L., et al. "Micromagnetic simulations using graphics processing units." Journal of Physics D: Applied Physics 45.32 (2012): 323001.

[8] Gregory, Kate, and Ade Miller. C++ AMP: Accelerated Massive Parallelism with Microsoft® Visual C++®. " O'Reilly Media, Inc.", 2012.

[9] Gilbert, Thomas L. "A phenomenological theory of damping in ferromagnetic materials." Magnetics, IEEE Transactions on 40.6 (2004): 3443-3449.

[10] Nakatani, Yoshinobu, Yasutaro Uesaka, and Nobuo Hayashi. "Direct solution of the Landau-Lifshitz-Gilbert equation for micromagnetics." Japanese Journal of Applied Physics 28.12R (1989): 2485.

[11] Hayashi, Nobuo, Koji Saito, and Yoshinobu Nakatani. "Calculation of demagnetizing field distribution based on fast Fourier transform of convolution." Japanese journal of applied physics 35.12A (1996): 6065-6073.

[12] Moth, Daniel, et al. "C++ AMP FFT Library." CodePlex, Jan 2013. Web. 23 Jun. 2014

[13] Michael Donahue, et al. "µMAG Standard Problem #3." µMAG organization. Mar 1998. Web. 23 Jun. 2014

**Append. A.** Comparison between APIs of OpenCL and C++ AMP.

| **OpenCL** | **C++ AMP** |
|---|---|
| size_t localCalculationSize[] = {32, 32};<br><br>size_t globalCalculationSize[] = {size, size};<br><br>err = clEnqueueNDRangeKernel(commandQueue, kernel, 2, NULL, globalCalculationSize, localCalculationSize, 0, NULL, NULL); | parallel_for_each(d_C.extent.tile<32, 32>(), |